\documentclass[twocolumn,showpacs,amsmath,amssymb]{revtex4}

\usepackage{graphicx}
\usepackage{bm}

\newcommand{\twovec}[2]
{\left( \begin{array}{c} #1 \\ #2 \end{array} \right)}

\begin{document}

\title{Dirac-field model of inflation in Einstein-Cartan theory}

\author{Tomoki Watanabe}

\email{tomoki@tokai-u.jp}

\affiliation{Department of Physics, Tokai University,\\
1117 Kitakaname, Hiratsuka, Kanagawa 259-1292, Japan}

\date{\today}

\begin{abstract}
We present a cosmological model in which a single Dirac field with a flat
potential can give rise to inflation within the framework of the
Einstein-Cartan theory.
It is shown that our Dirac-field model leads to a nearly scale-invariant
spectrum of density fluctuations owing to the spin-interaction which naturally
arises from the field equations of the Einstein-Cartan theory.
\end{abstract}


\maketitle

\section{Introduction}

Nowadays, it is widely accepted that inflation of the early universe
\cite{inflation} is due to one or more slow-rolling scalar fields,
called inflaton, which typically lead to a nearly scale-invariant power
spectrum of primordial density fluctuations with the spectral index
$n \approx 1$ in good agreement with recent, increasingly accurate
observations of cosmic microwave background anisotropy
\cite{wmap, wmap2, wmap3, wmap4}.
However, it remains a fundamental task to identify the inflaton with
a certain particle in promising particle-physics models theoretically
and experimentally.
Then the question arises: cannot other fields drive inflation?
Particularly we focus on Dirac spinor fields.

As the scalar field is often necessary in various theories of nature,
so the Dirac field is essential to a satisfactory description of relativistic
fermions.
Thus, it is of considerable importance and interest to investigate roles
of the Dirac field in cosmology.

On this subject, much remarkable work has been done within general relativity
by various authors.
For example, quantization of a Dirac field and studies of particle creation
in the expanding Friedmann-Robertson-Walker (FRW) universe were presented
in Refs.~\cite{quantum, quantum2, quantum3};
quantization of the system where a Dirac field couples to the FRW metric
was discussed in Refs.~\cite{isham, christ};
the explicit solutions of the Dirac equation in the FRW universe were derived
in Refs.~\cite{barut, kovalyov}.

Moreover, as for the issue of the accelerating universe, it was shown that
a self-interacting or nonlinear Dirac field can yield negative pressure
and thereby accelerate the early and the late-time expansion of the universe
\cite{ribas, saha}.
However, when we discuss whether a Dirac field can be an alternative to the
inflaton, we should pay attention to Armend{\' a}riz-Pic{\' o}n and Greene's
work \cite{picon} which concluded that, although the de Sitter expansion can
be driven by a flat potential of a Dirac field, such models lead to a
scale-dependent spectrum with $n = 4$, and thus are incompatible with the
observations.
As other ways not to rely on the inflaton, one can introduce vector fields
\cite{ford, golovnev}, or non-Dirac spinors that have some unusual properties
\cite{bohmer}.

On the other hand, since Dirac fields have spin-1/2 in contrast to scalar
fields, it is also of great interest to study cosmological effects of the
spin.
Then, one needs gravitational theories that naturally bring spin of matter
fields to the geometry of spacetime, since in general relativity microscopic
quantities such as the spin are usually neglected.
For this purpose, as one of such extended theories, we will adopt the
Einstein-Cartan theory \cite{hehl} in which nonvanishing torsion is
algebraically equivalent to spin of matter fields through the field equation
and, as a result, a spin-interaction is generated.
Generally, one can suppose affine connections in spacetime to be asymmetric,
thereby naturally obtaining a modified theory of general relativity.
Since, from the viewpoint of high energy physics, general relativity should
appropriately be extended so as to be able to describe the very early universe,
the Einstein-Cartan theory can be a reasonable framework for considering
inflation.
We can then expect that the spin has a crucial role in inflationary cosmology.

Many authors have also presented a variety of cosmological models based on
non-Riemannian gravity \cite{hehl2,hammond}, as exhaustively listed in
Ref.~\cite{dirk}.
The subject of early investigations based on the Einstein-Cartan theory was
mainly to construct exact solutions and to avoid the initial singularity
in the presence of torsion \cite{dirk, singularity, singularity3}.
Subsequently, after the advent of inflationary cosmology, inflationary
solutions were obtained also in the Einstein-Cartan models, some of which
\cite{gasperini, fennelly, assad, obukhov} relied on spin effects of the
so-called spinning fluid \cite{ray, mohseni, szydlowski}.

In this paper, we show that a single Dirac field can give rise to inflation
within the Einstein-Cartan theory, and prove compatibility of the Dirac-field
model with the observations by calculating the power spectrum of density
fluctuations of the Dirac field.

We will take the Dirac Lagrangian as the source of metric and torsion because
we are only concerned with seeking the origin of inflation in the context of
particle-physics.
It should be noted that we will deal with the homogeneous Dirac field
classically, namely, as a set of complex-valued functions that transform
according to the spinor representation of the Lorentz group and fulfill
the Dirac equation.
We here consider the expectation value of a spinor operator-valued field to
be the corresponding classical spinor;
a possible justification for the existence of classical spinors is made in
Ref.~\cite{picon}.
In addition to the fact that the classical spinors are mathematically
consistent, as shown in Ref.~\cite{picon}, such classical treatment for
a physical state can be a good approximation in some cases.

The organization of this paper is as follows:
in section \ref{sec2}, we briefly review the description of a Dirac field
coupled to gravity in the Einstein-Cartan theory and derive cosmological
equations in the flat FRW universe.
Within this framework, in section \ref{sec3}, after showing that a flat
potential of the Dirac field leads to the de Sitter expansion of the universe,
we compute the power spectrum of density fluctuations of the Dirac field.
We will observe that the obtained spectrum is nearly scale-invariant owing
to the existence of the spin-interaction.
We present our conclusion in section \ref{sec4}.

\section{\label{sec2}Formalism}

In this section, we summarize how the Einstein-Cartan theory describes
interactions between a Dirac field and gravity, and present the cosmological
equations that form the basis of our Dirac-field model of inflation.

In order to deal with a Dirac field $\psi$ coupled to gravity, one requires
that the action of the Dirac field be invariant under local Lorentz
transformations, with the help of the vierbein formalism
\cite{vierbein, vierbein2} in which vierbeins $e_\mu^{\;\;a}$ and spin
connections $\omega_\mu^{\;\; ab}$ are introduced as the fundamental
variables.
The vierbein satisfies $\eta_{ab} e_\mu^{\;\;a}e_\nu^{\;\;b} = g_{\mu\nu}$
with the Minkowski metric $\eta_{ab}$, where the signature we use is
$\eta_{ab} = \mathrm{diag} (+1, -1, -1, -1)$.
(The Greek and Latin indices denote spacetime and Lorentz indices
respectively.)
The spin connection $\omega_\mu^{\;\; ab}$ is defined by
$\omega_\mu^{\;\; ab} \equiv e^{\nu [a} \nabla_\mu e_\nu^{\;\;b]}$,
where $\nabla_\mu$ is a covariant derivative based on affine connections
of spacetime $\Gamma^\rho_{\;\;\mu\nu}$ and acts on tensors.
For the Dirac field, the covariant derivative based on the spin connection
is defined by
\begin{equation}
D_\mu \psi \equiv \partial_\mu \psi - \frac{i}{4} \omega_\mu ^{\;\;ab}
\sigma_{ab} \psi,
\label{cov_deriv}
\end{equation}
where $\sigma^{ab} \equiv (i/2) [ \gamma^a , \gamma^b ]$ is the generator
of the spinor representation of the Lorentz group and the constant
$\gamma$-matrices $\gamma^a$ satisfy the Clifford algebra
$\{\gamma^a, \gamma^b\} = 2 \eta^{ab}$;
the covariant derivative acting on the Dirac adjoint
$\bar{\psi} \equiv \psi^\dagger \gamma^{0}$ is
\begin{equation}
D_\mu \bar{\psi} \equiv
\partial_\mu \bar{\psi} + \frac{i}{4} \omega_\mu^{\;\;ab}
\bar{\psi} \sigma_{ab}.
\end{equation}
In this paper, for explicit calculations, we will choose the Dirac
representation
\begin{equation}
\gamma^0 =
\left(
\begin{array}{cc}
1 & 0 \\
0 & -1
\end{array}
\right),
\qquad
\gamma^m =
\left(
\begin{array}{cc}
0 & \sigma^m \\
-\sigma^m & 0
\end{array}
\right),
\end{equation}
where $\sigma^m$ are the conventional $2 \times 2$ Pauli matrices.
For later convenience, we also define the additional $\gamma$-matrix as
$\gamma_5 \equiv i \gamma^0  \gamma^1  \gamma^2  \gamma^3$.
It is obvious that the covariant derivative defined by Eq.~(\ref{cov_deriv})
transforms as a vector under diffeomorphisms and as a spinor under local
Lorentz transformations.
The coordinate components of $\gamma^a$ are defined by
\begin{equation}
\gamma^\mu \equiv e^\mu _{\;\;a} \gamma^a,
\end{equation}
which are shown to give a new set of $\gamma$-matrices that satisfy the
algebra $\{\gamma^\mu, \gamma^\nu\} = 2g^{\mu\nu}$.

Thus, the Dirac Lagrangian generalized into a curved spacetime background
is given by
\begin{equation}
\mathcal{L}_\psi =
\frac{i}{2}
\left[
\bar{\psi} \gamma^\mu D_\mu \psi - (D_\mu \bar{\psi}) \gamma^\mu \psi
\right] - V,
\label{dirac lagrangian}
\end{equation}
where the term $V$ generically represents a scalar potential of the Dirac
field including a mass term and self-interactions, and consists of arbitrary
functions of invariants generated from $\psi$ and $\bar{\psi}$;
in what follows, we assume $V = V (s)$ with $s \equiv \bar{ \psi} \psi$
for simplicity.

On the other hand, for gravity, we take the Einstein-Hilbert Lagrangian
\begin{equation}
\mathcal{L}_g =
- \frac{1}{16\pi G} e^\mu_{\ a} e^\nu_{\ b} R^{ab}_{\ \ \mu\nu},
\end{equation}
where $R^{ab}_{\ \ \mu\nu}$ is the curvature of the spacetime given by
\begin{equation}
R^{ab}_{\ \ \mu\nu} =
\partial_\mu \omega^{\;\;ab}_\nu -\partial_\nu \omega^{\;\;ab}_\mu
+ \omega^{\;\;ac}_\mu \omega^{\ \ \> b}_{\nu c}
- \omega^{\;\;ac}_\nu \omega^{\ \ \> b}_{\mu c}.
\label{R omega}
\end{equation}

In order to take into account effects of the spin of a Dirac field on gravity,
it is necessary to extend general relativity.
In this paper, for this purpose, we introduce the Einstein-Cartan theory
\cite{hehl} in which nonvanishing torsion
$C^\rho_{\;\;\mu\nu}\equiv 2 \Gamma^\rho_{\;\;[\mu\nu]}$
is related with spin of matter fields.
(For a very brief review of the Einstein-Cartan theory, see
Appendix \ref{app}.)

Within this framework, we consider a system in which a Dirac field provides
the unique source of metric and torsion of spacetime.
Since the existence of torsion leads to the spin-interaction in the curvature
(\ref{R omega}) and the kinetic terms of the Lagrangian
(\ref{dirac lagrangian}), the conventional Einstein equation is modified as
\begin{equation}
\tilde{G}_{\mu\nu} =
8\pi G
\left(
\tilde{T}_{(\mu\nu)} - \frac{3 \pi G }{2} \phi_a \phi^a g_{\mu\nu}
\right)
\equiv
8 \pi G T_{\mu\nu}^\mathrm{(tot)}.
\label{E eq for Dirac}
\end{equation}
The additional term $- 3 \pi G \phi_a \phi^a g_{\mu\nu}/2$ with
$\phi^a \equiv \bar{\psi} \gamma_5 \gamma^a \psi$ is the spin term prescribed
by the Einstein-Cartan theory, and can be regarded as a correction due to
the spin of the Dirac field.
Here, the tilde indicates quantities free from torsion;
$\tilde{G}_{\mu\nu}$ is the Einstein tensor composed of the Riemannian
(Levi-Civita) connection;
$\tilde{T}_{(\mu\nu)}$ is the usual, symmetric energy-momentum tensor of
the Dirac field, given by
\begin{equation}
\tilde{T}_{(\mu\nu)} =
\frac{i}{2}
\left[
\bar{\psi} \gamma_{(\mu} \tilde{D}_{\nu)} \psi
-(\tilde{D}_{(\nu} \bar{\psi}) \gamma_{\mu)} \psi
\right]
- g_{\mu\nu} \tilde{\mathcal{L}}_\psi,
\end{equation}
where $\tilde{D}_{\mu}$ is the covariant derivative based on the Riemannian
spin connection
$\tilde{\omega}_\mu^{\;\; ab}
\equiv e^{\nu [a} \tilde{\nabla}_\mu e_\nu^{\;\;b]}$,
and $\tilde{\mathcal{L}}_\psi$ is the torsion-free Lagrangian of the Dirac
field defined by
\begin{equation}
\tilde{\mathcal{L}}_\psi \equiv \frac{i}{2}
\left[
\bar{\psi} \gamma^\mu \tilde{D}_\mu \psi
- (\tilde{D}_\mu \bar{\psi}) \gamma^\mu \psi
\right] - V.
\end{equation}
It is interesting that, in Eq.~(\ref{E eq for Dirac}), the spin correction
to the energy-momentum tensor of the Dirac field is proportional to both
the metric $g_{\mu\nu}$ and the gravitational constant $G$;
such a simple form strongly motivates us to study the Dirac-gravity system
including effects of the spin.

Similarly, the conventional Dirac equation in curved spacetime is also
modified so as to include the spin-interaction term:
\begin{equation}
i \gamma^\mu \tilde{D}_\mu \psi - V' \psi
+ 3 \pi G \phi_a \gamma_5 \gamma^a \psi = 0,
\label{general eom in EC}
\end{equation}
where the prime denotes the derivative with respect to $s$.
As can be seen from Eqs.~(\ref{E eq for Dirac}) and (\ref{general eom in EC}),
one can arrive at the usual general-relativistic equations whenever
the energy scale of the spin-interaction $G \phi_a \phi^a$ is negligible
in comparison with typical energy scales of the kinetic term or the potential.

With the general formalism described above, we are now interested to
investigate cosmology.
Let us consider the flat FRW universe, in which the metric is given by
\begin{equation}
ds^2 = dt^2 - a^2(t) d\mathbf{x}^2,
\end{equation}
and the vierbein is chosen to be
\begin{equation}
(e_\mu^{\ a})
= \mathrm{diag} [1, a(t), a(t), a(t)],
\end{equation}
where $t$ is the cosmic time and $a (t)$ is the scale factor.
(See Refs.~\cite{isham, christ} for the nonflat FRW universe.)
Then we should exclude any spatial dependence of the Dirac field for
consistency with homogeneity of the spacetime: $\partial_i \psi = 0$.
Consequently, we obtain the following Dirac equation:
\begin{equation}
\dot{\psi} + \frac{3}{2} H \psi + i \gamma^0 V' \psi
- 3 i \pi G \phi_a \gamma^0 \gamma_5 \gamma^a \psi = 0,
\label{eom in EC}
\end{equation}
where $H \equiv \dot{a}/a$ is the Hubble parameter and the dot denotes the
derivative with respect to $t$.
From Eq.~(\ref{eom in EC}), the anisotropic components $\tilde{T}_{(0i)}$
can be shown to vanish, which is consistent with isotropy of the spacetime,
$\tilde{G}_{0i} = 0$.

Now the energy density and pressure, which must be specified to describe
cosmological dynamics, can be found from the total energy-momentum tensor
defined in Eq.~(\ref{E eq for Dirac}) as
\begin{eqnarray}
\rho_\mathrm{tot} &\equiv& 
{}^\mathrm{(tot)} T_0 ^{\;\;0} = V -  \frac{3 \pi G }{2} \phi_a \phi^a,
\\
p_\mathrm{tot} &\equiv& 
- {}^\mathrm{(tot)} T_i ^{\;\;i} =
s V' - V - \frac{3 \pi G }{2} \phi_a \phi^a.
\label{pressure}
\end{eqnarray}
It should be noted that $\phi_a \phi^a$ is always negative, and consequently
that the spin component in Eq.~(\ref{pressure}) contributes as extra positive
pressure.
In terms of these variables, the cosmological evolution equations are given by
\begin{eqnarray}
H^2 &=&
\frac{8 \pi G}{3} \left( V -  \frac{3 \pi G }{2} \phi_a \phi^a \right),
\\
\frac{\ddot{a}}{a} &=&
- \frac{4 \pi G}{3}
\left(
- 2 V + 3 s V' - 6 \pi G \phi_a \phi^a
\right),
\label{de for a}
\end{eqnarray}
which are linked through the conservation law
\begin{equation}
\dot{\rho}_\mathrm{tot} + 3 H ( \rho_\mathrm{tot} + p_\mathrm{tot} ) = 0.
\label{conservation}
\end{equation}
One can verify that the conservation law (\ref{conservation}) is equivalent
to the Dirac equation (\ref{eom in EC}).
Therefore, the system that consists of a Dirac field and the flat FRW
spacetime is self-consistent in the Einstein-Cartan theory as well as
in general relativity.

Finally, we define the equation of state $w$ as
\begin{equation}
w \equiv \frac{ p_\mathrm{tot} }{ \rho_\mathrm{tot} } =
\frac{s V' - V - 3 \pi G \phi_a \phi^a/2}
{V - 3 \pi G \phi_a \phi^a/2}.
\label{eos in EC}
\end{equation}
It is worth mentioning that, by taking the torsion-free limit
$G \phi_a \phi^a / V \to 0$, we can always reproduce the cosmological
equations in general relativity formulated by Armend{\' a}riz-Pic{\' o}n and
Greene \cite{picon}.
As an important example, the equation of state they found is
\begin{equation}
w_\mathrm{GR} = s \frac{V'}{V} - 1.
\label{eos in GR}
\end{equation}

\section{\label{sec3}Inflation}
\subsection{Background}

In this subsection, we show that a Dirac field can cause the de Sitter
expansion of the background universe on the basis of the cosmological
equations collected in the previous section.

In general relativity, from Eq.~(\ref{eos in GR}), one can discuss de Sitter
inflation driven by a Dirac field, as done in Ref.~\cite{picon}.
In this case, $V$ is assumed to be sufficiently flat because
$w_\mathrm{GR} \approx -1$ is guaranteed by
\begin{equation}
\left| \frac{d \ln V}{d \ln s} \right| \ll 1,
\label{slow-roll}
\end{equation}
which is a similar requirement to the slow-roll condition for conventional
scalar-field models of inflation.
The condition (\ref{slow-roll}) is satisfied if $V$ is asymptotic to a
positive constant for large $s$.
For example, $\ln (1 + s^n)$, $\tanh (n s)$, and $s^n/(1 + s)^n$ with positive
$n$ are candidates for a potential that realizes the de Sitter expansion.

Also in our case, the flatness of $V$ is a basic premise for considering
inflation.
First, in the evolution equation for $\phi_a \phi^a$,
\begin{equation}
\frac{d}{dt}\left( \phi_a \phi^a \right) + 6 H \phi_a \phi^a
+ 4 i V' \phi^0 \bar{\psi} \gamma_5 \psi = 0,
\label{eom for phi phi}
\end{equation}
which is derived from Eq.~(\ref{eom in EC}), the flatness condition allows us
to ignore the third term on the left hand side and hence leads to
$\phi_a \phi^a \propto a^{-6}$.
Then it follows that all the spin components, which are incorporated in the
form of $G \phi_a \phi^a$ in the background equations presented previously,
decrease faster than the other terms in those equations as the universe
expands.

\begin{figure}[htbp]
\includegraphics{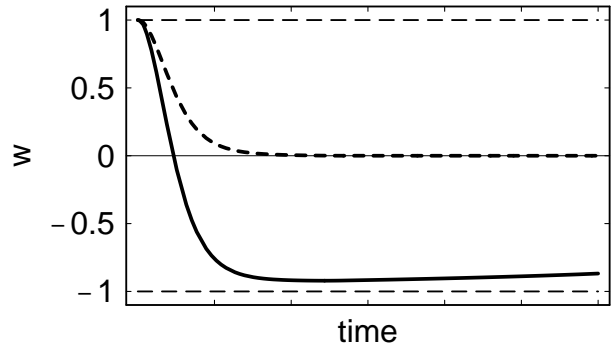}
\caption{\label{figure1}
Typical time evolution of the equation of state $w$.
The solid curve represents $w$ for the solution of the cosmological
equations (\ref{eom in EC}) and (\ref{de for a}) with the potential
of the form $\ln (1 + s^2)$,
and shows that its value which is initially $w \approx 1$ comes close
to $-1$ as time progresses.
The dashed curve represents the simultaneous evolution of the spin
components in $w$,
$(- 3 \pi G \phi_a \phi^a /2) (V - 3 \pi G \phi_a \phi^a/2)^{-1}$,
disappearing with time.
}
\end{figure}

When the spin components become so small as to be negligible, all the
background equations in the previous section arrive back at those of
Ref.~\cite{picon}.
If the potential retains the sufficient flatness until that time, then
the equation of state eventually becomes
$w \approx w_\mathrm{GR} \approx -1$.
Thus, a flat potential guarantees de Sitter inflation in our model as well.
In FIG.\ \ref{figure1}, we have numerically demonstrated that $w$ approaches
the value of $-1$ as the spin components decrease under a sufficiently flat
potential.

Meanwhile, we can also speculate about the early universe before the
inflation era.
The result that the spin components evolve as $\phi_a \phi^a \propto a^{-6}$
also means that these become greater in earlier stages of the universe.
Thus it can be seen that the very early universe is dominated by the spin
components in our model, and also that, from Eq.~(\ref{eos in EC}), during
such an early epoch the Dirac field satisfies $w \approx 1$ which corresponds
to the equation of state for a stiff fluid.
Therefore, our model suggests that inflation begins when the spin-dominated
era ends, which can be seen also from FIG.\ \ref{figure1}.
Since the Einstein-Cartan theory is expected to be relevant to high energy
physics, especially theories of supergravity \cite{sugra, sugra2}
in which the algebraic equivalence between torsion and spin is necessary
for the construction of supersymmetric transformations, it is acceptable that
the spin components dominate the very early universe which should be considered
to be at high temperature.

On the other hand, the evolution of $s$ is given by $s \propto a^{-3}$ during
inflation because Eq.~(\ref{eom in EC}) can be recast as
\begin{equation}
\dot{s} + 3 H s + 6 i \pi G \phi^0 \bar{\psi} \gamma_5 \psi = 0,
\end{equation}
where the third term on the left hand side arises from the spin-interaction,
and hence approximately vanishes when inflation begins.
Consequently, number of $e$-foldings between a time $t_1$ and a later time
$t_2$ is given by
\begin{equation}
N = \frac{1}{3} \ln \frac{s(t_1)}{s(t_2)}.
\end{equation}
Therefore, if one requires $N \approx 60$ in order to solve the horizon
problem, then $s$ must change by approximately eighty orders of magnitude.
In this study we assume the potential to be so sufficiently flat as to satisfy
Eq.~(\ref{slow-roll}) for such a wide range of $s$ regardless of its origin;
our purpose is to investigate whether the Dirac field with a flat potential
can be a source of inflation in the Einstein-Cartan theory.

Similarly to the scalar $s$, the pseudoscalar $s_5$ defined by
$s_5 \equiv \bar{\psi} \gamma_5 \psi$ evolves according to
$s_5 \propto a^{-3}$ during inflation because $s_5$ fulfills the evolution
equation
\begin{equation}
\dot{s}_5 + 3 H s_5 + 2 i V' \phi^0 + 6 i \pi G \phi^0 s = 0,
\end{equation}
where both the last and the penultimate terms on the left hand side are
negligible under our assumption.
We also note that, in terms of the rescaled field defined by
$\Psi \equiv a^{3/2} \psi$, these scalars, $s$ and $s_5$, can be rewritten as
$s = a^{-3} \bar{\Psi} \Psi$, $s_5 = a^{-3} \bar{\Psi} \gamma_5 \Psi$
respectively, where $\bar{\Psi} \Psi$ and $\bar{\Psi} \gamma_5 \Psi$ are
constant as long as the sufficient flatness of the potential holds.

\subsection{Perturbation}

In this subsection, we discuss perturbations of the Dirac field
that brings about de Sitter inflation via the dynamics explained above.

The conventional scalar-field models of inflation \cite{inflation} have an
important feature of predicting a nearly scale-invariant spectrum of density
fluctuations in excellent agreement with recent observations
\cite{wmap, wmap2, wmap3, wmap4}.
Our next task is thus to examine the consistency of the Dirac-field model of
inflation with the observations, namely, whether our model can derive a nearly
scale-invariant spectrum, by computing the power spectrum of density
perturbations of the Dirac field.
Although a proper analysis for this purpose should be based on gauge-invariant
perturbation theories \cite{weinberg, kodama, sasaki, mukhanov} where both
spacetime and matter fields are perturbed, we will not consider the metric
perturbations but only the perturbed field $\delta \psi$ for simplicity.

Our key strategy is to perturb the spin components of the background
equations.
Whereas Armend{\' a}riz-Pic{\' o}n and Greene \cite{picon} showed that,
within the framework of general relativity, Dirac-field models of inflation
with a flat potential lead to a scale-dependent spectrum and hence are
inconsistent with the observations, our model based on the Einstein-Cartan
theory possesses in the first place the spin-interaction, which opens the
possibility of improving their result.
It should be noted that the terms related to the spin in our model appear as
a consequence of a natural extension of general relativity.
As discussed previously, the inflationary expansion itself can be shown
to occur by simply assuming the flat potential, whether the Dirac field has
the spin-interaction or not;
however, if the spin terms exist, then their fluctuations must also exist.
We will show below that our model predicts a nearly scale-invariant spectrum
owing to these fluctuations.

In what follows, we mark the background quantities with $B$, and for
convenience, use the identity
\begin{equation}
\phi_a \phi^a = - (\bar{\psi} \psi)^2 - (i \bar{\psi} \gamma_5 \psi)^2
\equiv - s^2 + s_5^2.
\end{equation}
The scalar $s = \bar{\psi} \psi$ and the pseudoscalar
$s_5 = \bar{\psi} \gamma_5 \psi$ are now perturbed as
\begin{eqnarray}
s &=&
\bar{\psi}_B \psi_B + \bar{\psi}_B \delta \psi + \delta \bar{\psi} \psi_B
\equiv s_B + \delta s,
\\
s_5 &=&
\bar{\psi}_B \gamma_5 \psi_B
+ \bar{\psi}_B \gamma_5 \delta \psi + \delta \bar{\psi} \gamma_5 \psi_B
\equiv s_5^B + \delta s_5.
\end{eqnarray}
The equation of motion for $\delta \psi$ is then given by
\begin{eqnarray}
\lefteqn{
i \gamma^0 \left( \delta \dot{\psi} + \frac{3}{2} H \delta \psi \right)
+ i \frac{1}{a} \gamma^m \partial_m \delta \psi
- m \delta \psi
}
\nonumber \\
&& \!\!\!
- 3 \pi G
\left[
( s_B - s_5^B \gamma_5 ) \delta \psi
+ ( \delta s - \delta s_5 \gamma_5 ) \psi_B
\right]
= 0,
\label{eom for delta psi}
\end{eqnarray}
where $m \equiv V'$ and we have ignored $V''$ because of the flatness of $V$,
from which it follows that $m$ is approximately constant.
Again we note that the term proportional to $G$ is the correction due to
torsion or equivalently spin;
i.e., one can take the limit $G \to 0$ as a method to render the perturbed
equations torsion-free in the dynamics of perturbations, while, in the
background equations, the torsion-free equations are reproduced as a result
of rapid disappearance of the spin components of the form $G \phi_a \phi^a$.
On the other hand, taking account of the fact that general relativity provides
a successful description of the present universe, one can consider the
correction coming from torsion to be generally small.
In other words, we can regard the term proportional to $G$ as a first-order
correction to general relativity, and will ignore higher-order corrections in
$G$ whenever the quantities derived below include such terms.

In the limit $G \to 0$, the torsion-free equation of motion is found to be
\begin{equation}
i \gamma^\mu \tilde{D}_\mu \delta \tilde{\psi} - m \delta \tilde{\psi} = 0,
\label{eom for delta psi in GR}
\end{equation}
which can be solved analytically \cite{barut}.
For $H \approx \mathrm{const.}$, the plane-wave solution $u_s (\mathbf{k},t)$
of Eq.~(\ref{eom for delta psi in GR}) is generally given by
\begin{equation}
u_s =
a^{-3/2} \sqrt{ \frac{- \pi k \eta }{2} }
\twovec{
\alpha_k^{+} H_{\nu}^{(1)} (- k \eta)
+ \beta_k^{+} H_{\nu}^{(2)} (- k \eta)
}
{
\alpha_k^{-} H_{\nu^*}^{(1)} (- k \eta)
+ \beta_k^{-} H_{\nu^*}^{(2)} (- k \eta)
},
\label{general sol in GR}
\end{equation}
where $\eta$ is the conformal time and $H_{\nu}^{(1,2)} (-k\eta)$ are
the Hankel functions of the first and the second kind with
$\nu = 1/2 - i m/H \approx \mathrm{const.}$;
$\alpha_k^{\pm}$ and $\beta_k^{\pm}$ are arbitrary constant two-spinors.
The solution of Eq.~(\ref{eom for delta psi in GR}) can then be expanded as
\begin{eqnarray}
\delta \tilde{\psi} &=&
\int \frac{d^3 k}{(2\pi)^{3/2}} \sum_{s=1}^2
\big[
u_s (\mathbf{k},t) a_s (\mathbf{k}) e^{i \mathbf{k} \cdot \mathbf{x}}
\nonumber \\
&& \quad
+ v_s (\mathbf{k},t) b_s^\dag (\mathbf{k}) e^{- i \mathbf{k} \cdot \mathbf{x}}
\big],
\end{eqnarray}
where the mode function $v_s (\mathbf{k},t)$ corresponding to negative energy
solutions is of the same form as Eq.~(\ref{general sol in GR}).

Now, for the purpose of investigating the changes due to the spin from the
torsion-free case, it is useful to introduce scalar functions
$A(\mathbf{k},t)$ and $B(\mathbf{k},t)$,
and to express the plane-wave expansion of the solution of
Eq.~(\ref{eom for delta psi}) in terms of the analytical solution
(\ref{general sol in GR}) as
\begin{eqnarray}
\delta \psi &=&
\int \frac{d^3 k}{(2\pi)^{3/2}}
\sum_{s = 1}^2
\big[
A(\mathbf{k},t) u_s (\mathbf{k}, t)
a_s (\mathbf{k}) e^{i \mathbf{k} \cdot \mathbf{x}}
\nonumber \\
&& \quad
+ B(\mathbf{k},t) v_s (\mathbf{k}, t)
b_s^\dag (\mathbf{k}) e^{- i \mathbf{k} \cdot \mathbf{x}}
\big],
\label{delta psi}
\end{eqnarray}
in which expression, $A$ and $B$ play a role to represent the modifications
to the torsion-free case; i.e., by substituting Eq.~(\ref{delta psi})
into Eq.~(\ref{eom for delta psi}), one can see that
the evolution of $A$ and $B$ is governed by the spin-interaction part
of Eq.~(\ref{eom for delta psi}), and also that both $A$ and $B$ are constant
in the limit $G \to 0$.
Here, $a_s$ and $b_s$ are the particle and antiparticle operators satisfying
$
\{ a_s (\mathbf{k}), a_{s'}^\dag (\mathbf{k}') \} =
\{ b_s (\mathbf{k}), b_{s'}^\dag (\mathbf{k}') \} =
\delta_{s s'} \delta^3 (\mathbf{k} - \mathbf{k}')
$
under an appropriate orthonormality.
Then, the vacuum expectation value of the square of $\delta \psi$ is
\begin{equation}
\langle \delta \bar \psi \delta \psi \rangle
= \int \frac{d^3 k}{(2\pi)^{3}}
|B|^2 \sum_s \bar{v}_s (\mathbf{k}) v_s (\mathbf{k}).
\end{equation}

Next, we characterize density perturbations $\delta \rho$ by the variable
\begin{equation}
\zeta \equiv \frac{\delta \rho}{\rho + p},
\end{equation}
This quantity is the one employed in Ref.~\cite{picon}, defined analogously
to the gauge-invariant Bardeen variable, which is conserved for adiabatic
perturbations on sufficiently large scales \cite{weinberg, wands}.
For our model, we have $\rho + p = m s_B$ and
\begin{equation}
\delta \rho = m \delta s + 3 \pi G (s_B \delta s - s_5^B \delta s_5),
\end{equation}
from which we compute the power spectrum $\mathcal{P}$ of the variable $\zeta$
on comoving scales $1/k$, defined by
\begin{equation}
\langle \zeta (\mathbf{x}, t) \zeta (\mathbf{x} + \mathbf{r}, t) \rangle
= \int \frac{dk}{k} \frac{\sin kr}{kr} \mathcal{P}(k).
\end{equation}
We have supposed that the Fourier transform of the correlation function on
the left hand side depends only on $k$ and not on $\mathbf{k}$ itself,
for consistency with the isotropy of the background spacetime.

In order to evaluate the isotropic power spectrum of
$\langle \zeta^2 \rangle$, following the method of Ref.~\cite{picon}, we apply
the Fierz transformation to $\langle \zeta^2 \rangle$ and expand the vacuum
expectation values of the quadratic term of $\delta s$ and $\delta s_5$ in
terms of perturbation bilinears.
For example, the vacuum expectation value of $\delta s^2$ can be expressed as
\begin{equation}
\langle \delta s^2 \rangle =
\frac{s_B}{2} \langle \delta \bar{\psi} \delta \psi \rangle
+ \frac{ \bar{\psi}_B \gamma^a \psi_B }{2}
\langle \delta \bar{\psi} \gamma_a \delta \psi \rangle + \cdots.
\end{equation}
After that, as noted in Ref.~\cite{picon}, we concentrate on the perturbations
of the scalar bilinear form $\langle \delta \bar{\psi} \delta \psi \rangle$
and discard the remaining vectorial bilinears that may break isotropy of the
power spectrum because we are only concerned with determining the amplitude
and $k$-dependence of $\mathcal{P}$.
In this way, $\mathcal{P}$ is obtained as
\begin{equation}
\mathcal{P}(k)
= \frac{|B|^2 k^3}{4\pi^2 \bar{\Psi}_B \Psi_B}
\left(
1 + \frac{3 \pi G C_1 a^{-3}}{m}
\right)
\sum_s \bar{V}_s V_s,
\label{P}
\end{equation}
where $V_s$ is the rescaled mode function defined by $V_s \equiv a^{3/2} v_s$
and
$
C_1 = 2 \bar{\Psi}_B \Psi_B
- (\bar{\Psi}_B \gamma_5 \Psi_B)^2/\bar{\Psi}_B \Psi_B
$
is a constant.
We have here ignored the terms of $O(G^2)$ as mentioned before.
In comparison with the result of Ref.~\cite{picon}, the extra factor $|B|^2$
appears in addition to the first-order correction in the parentheses
in the above equation.
The effective part of $B$ that contributes to the isotropic power spectrum
fulfills the following equation:
\begin{eqnarray}
\frac{\dot{B}}{B}
&=& - \frac{3 i \pi G}{\sum_s \bar{V}_s V_s}
\sum_s \big( s_B \bar{V}_s \gamma^0 V_s
- s_5^B \bar{V}_s \gamma^0 \gamma_5 V_s
\nonumber \\
&& \quad
+ \bar{\psi}_B V_s \bar{V}_s \gamma^0 \psi_B
- \bar{\psi}_B \gamma_5 V_s \bar{V}_s \gamma^0 \gamma_5 \psi_B
\big),
\label{eom for B}
\end{eqnarray}
which is obtained by multiplying Eq.~(\ref{eom for delta psi}) by
$\delta \bar{\psi}$ with Eq.~(\ref{delta psi}) and evaluating in a vacuum.

Now we have to determine the mode function $V_s$ and $B$.
We here take the Bunch-Davies vacuum as an appropriate initial vacuum state
\cite{vierbein2}, and choose the constants in $V_s$ so that $V_s$ coincides
with the Minkowski solution $V_s \propto e^{ik\eta}$ on sufficiently short
scales $k \to \infty$:
\begin{equation}
V_s =
\frac{ \sqrt{ - \pi k \eta } }{2}
\twovec{
e^{ - \pi m/2H } H_{\nu}^{(2)} (- k \eta )
\frac{ \mathbf{k} \cdot \bm{\sigma} }{k} \omega^s
}
{
e^{ \pi m/2H } H_{\nu^*}^{(2)} (- k \eta ) \omega^s
},
\end{equation}
where $\omega^s$ is the constant two-spinor defined by
$\omega^s_\alpha = \delta^s_\alpha$.
With this choice, in order to solve Eq.~(\ref{eom for B}), we use the
explicit background solutions during inflation,
\begin{eqnarray}
a &=& a_* e^{H t}
\\
\Psi_B &=& \twovec{\varphi_0 e^{- i m t}}{\chi_0 e^{i m t}},
\end{eqnarray}
where $\varphi_0$ and $\chi_0$ are arbitrary constants.
Moreover, fortunately, we can make use of the fact that $\sum_s \bar{V}_s V_s$
is approximately constant with respect to $z = - k \eta$.
Roughly speaking, the relation
$|d (\sum_s \bar{V}_s V_s)/dz| \ll |\sum_s \bar{V}_s V_s |$
holds for $z > 1$ and near $z = 0$ independently of $m$.
In fact, the asymptotic formulae of the Hankel functions \cite{abramowitz}
lead to
$\sum_s \bar{V}_s V_s \sim - 2 (m/H) z^{-1}$ for large $z$ and
$\sum_s \bar{V}_s V_s \sim -2 \tanh (\pi m/H)$ for small $z$.
We can safely employ this relation because, as discussed below, we only need
to find the asymptotic behavior of $|B|^2$ on large and small scales.
Then Eq.~(\ref{eom for B}) can be integrated to give
\begin{widetext}
\begin{eqnarray}
|B|^2
&\approx&
1 + \frac{3 i \pi^2 G H^2}{2 \sum_s \bar{V}_s V_s k^3}
\bigg\{
\left( \frac{k}{a_* H} \right)^{2 i m/H}
\frac{ \varphi_0^\dag \chi_0 z^3}{\cosh^2 \frac{\pi m}{H}}
\nonumber \\
&&
\times
\bigg[
- i \sinh \frac{\pi m}{H} \Gamma \left( \nu + \frac{3}{2} \right) z^{2\nu}
{}_2 F_3 \left( \frac{3}{2}, \nu + \frac{3}{2};
\nu + 1, \nu^* + 1, \nu + \frac{5}{2}; - z^2 \right)
\nonumber \\
&&
- \frac{ z^{- 4i m/H} }{ \sqrt{\pi} }
\Gamma \left( \nu + \frac{1}{2} \right)
\Gamma \left( 2 \nu + \frac{1}{2} \right)
{}_2 F_3 \left( \nu + \frac{1}{2}, 2\nu + \frac{1}{2};
\nu, 2\nu, 2\nu + \frac{3}{2}; - z^2 \right)
\nonumber \\
&&
+ \frac{1}{2} \Gamma \left( \nu^* + \frac{1}{2} \right)
{}_2 F_3 \left( \frac{3}{2}, \nu^* + \frac{1}{2};
\frac{5}{2}, \nu^*, 2 \nu^*; - z^2 \right)
\bigg]
- h.c. + C_2
\bigg\},
\label{B^2}
\end{eqnarray}
\end{widetext}
where ${}_2F_3 (a_1, a_2; b_1, b_2, b_3; x)$ is the regularized hypergeometric
function defined by
\begin{eqnarray}
\lefteqn{
{}_2 F_3 (a_1, a_2 ; b_1, b_2, b_3 ; x)
}
\nonumber \\
&\equiv&
\frac{1}{\Gamma(b_1) \Gamma(b_2) \Gamma(b_3)}
\sum_{k=0}^{\infty}
\frac{(a_1)_k (a_2)_k}{(b_1)_k (b_2)_k (b_3)_k}
\frac{x^k}{k !},
\end{eqnarray}
with the Pochhammer symbol $(a)_k \equiv \Gamma(a+k)/\Gamma(a)$,
and we have dropped the higher-order terms in $G$.
The integration constant $C_2$ should be chosen so that $|B|^2 \to 1$
on sufficiently short scales $k \to \infty$.
Using the asymptotic expansion of ${}_2F_3 (a_1, a_2; b_1, b_2, b_3; x)$
for $|x| \to \infty$,
\begin{eqnarray}
\lefteqn{
{}_2F_3 (a_1, a_2; b_1, b_2, b_3; x)
}
\nonumber \\
&\sim&
\frac{ (-x)^c }
{ 2 \sqrt{\pi} \Gamma(a_1) \Gamma(a_2)}
\left(
e^{- i ( \pi c + 2 \sqrt{-x} ) }
+ e^{ i ( \pi c + 2 \sqrt{-x} ) }
\right)
\nonumber \\
&&
+ \frac{\Gamma(a_2 - a_1) (-x)^{-a_1}}
{\Gamma(b_1 - a_1) \Gamma(b_2 - a_1) \Gamma(b_3 - a_1) \Gamma(a_2)}
\nonumber \\
&&
+ \frac{\Gamma(a_1 - a_2) (-x)^{-a_2}}
{\Gamma(b_1 - a_2) \Gamma(b_2 - a_2) \Gamma(b_3 - a_2) \Gamma(a_1)},
\end{eqnarray}
with $c = ( a_1 + a_2 - b_1 - b_2 - b_3 + 1/2 )/2$, we have
\begin{eqnarray}
C_2 &=& \mathrm{sech} \frac{2\pi m}{H}
\bigg[
\frac{ \varphi_0^\dag \chi_0 \Gamma\left( - \nu \right) }
{ \Gamma\left( -\nu - \frac{1}{2} \right)
\Gamma\left( \frac{1}{2} - 2 \nu \right) }
\left( \frac{k}{a_* H} \right)^{2 i m/H}
\nonumber \\
&&
- \frac{\chi_0^\dag \varphi_0 \Gamma\left( - \nu^* \right)}
{ \Gamma\left( -\nu^* - \frac{1}{2} \right)
\Gamma\left( \frac{1}{2} - 2 \nu^* \right) }
\left( \frac{k}{a_* H} \right)^{-2 i m/H}
\bigg].
\end{eqnarray}

The power spectrum $\mathcal{P}$ outside horizon is obtained by evaluating
Eq.~(\ref{P}) on large scales $k \to 0$ with Eq.~(\ref{B^2}), in which
we can use the asymptotic formula
$
{}_2F_3 (a_1,a_2; b_1,b_2,b_3; x)
\sim (\Gamma(b_1) \Gamma(b_2) \Gamma(b_3))^{-1}
$
for $x \to 0$.
Since the observed spectral index at horizon crossing is defined by
$n - 1 \equiv d \ln \mathcal{P}/d\ln k |_{k=aH}$, we subsequently estimate
$\mathcal{P}$ at $k = aH$.
For simplicity, let us here utilize an approximation $m/H \ll 1$, which is
allowed by the flatness of $V$.
Thus we find
\begin{widetext}
\begin{eqnarray}
\mathcal{P}|_{k = aH}
&\approx&
- \frac{G H^2}{\bar{\Psi}_B \Psi_B}
\bigglb(
\frac{3}{2} C_1
+ \frac{m}{H} (\varphi_0^\dag \chi_0 + \chi_0^\dag \varphi_0)
\left(
\frac{1}{3\pi} (3 \gamma + \ln 8 - 1) - \frac{3}{8}
\right)
\nonumber \\
&&
- \frac{m^2}{H^2}
\bigg\{
\frac{\pi^2}{2} C_1
+ i (\varphi_0^\dag \chi_0 - \chi_0^\dag \varphi_0)
\bigg[
\left(
\frac{3}{4} - \frac{2}{3\pi} (3 \gamma + \ln 8 - 1)
\right) \ln \left( \frac{k}{a_* H} \right)
\nonumber \\
&&
- \frac{2}{9\pi} (3 \gamma + \ln 8 - 2) + \frac{3}{8} (2 \gamma + \ln 4 - 3)
\bigg]
\bigg\}
\biggrb),
\end{eqnarray}
\end{widetext}
where $\gamma$ is Euler's constant, $\gamma \approx 0.577$.
Therefore, the spectral index is
\begin{equation}
n - 1 =
\frac{m^2}{H^2 C_1} i (\varphi_0^\dag \chi_0 - \chi_0^\dag \varphi_0)
\left(
\frac{4}{9\pi} (3 \gamma + \ln 8 - 1) - \frac{1}{2}
\right),
\end{equation}
which is our final result.
It should be emphasized that the $k$-dependence of $\mathcal{P}$ is naturally
suppressed by the condition $m/H \ll 1$ which follows from the flatness of
the potential, and hence that $n$ is nearly equal to 1.
We also note that the running index $\alpha = dn/d\ln k$ is further
suppressed, of the order $O(m^3)$.
Therefore, we are led to the conclusion that our Dirac-field model of
inflation can predict a nearly scale-invariant spectrum of density
fluctuations in agreement with the observations.

\section{\label{sec4}Summary}

We have shown that the Dirac-field model of inflation leads to a nearly
scale-invariant spectral index consistent with the observations,
by naturally extending the theoretical framework beyond general relativity.
It is usually believed that general relativity does not hold in the very early
universe and then should be extended appropriately to high energy physics;
the Einstein-Cartan theory adopted here is one of such extended theories.

In the framework of general relativity, Armend{\' a}riz-Pic{\' o}n and
Greene \cite{picon} found that a Dirac field with a flat potential can give
rise to the de Sitter expansion of the universe, but concluded that
the Dirac field itself cannot be an alternative to the conventional inflaton
field as the unique source of inflation because the spectral index obtained
from density fluctuations of the Dirac field, $n = 4$, is in strong
disagreement with the observations.
The key ingredient for improving their result, namely, obtaining
$n \approx 1$, is to introduce a spin-interaction, which naturally appears in
the Einstein-Cartan theory, into the dynamics of the Dirac field that has
the inflationary potential.

Because of the existence of the spin-interaction, the new terms of the form
$G \phi_a \phi^a$ must be added to both the cosmological background equations
and the equation of motion for the perturbed field presented in
Ref.~\cite{picon}.
However, we have seen that, according to Eq.~(\ref{eom for phi phi}),
the spin terms in the background equations decrease fast as the universe
expands; therefore, during inflation, we can ignore the effects due to the
spin on the background.
Without any change in the basic idea that a flat potential of a Dirac field
leads to an inflationary expansion, we have been able to gain the possibility
to solve the problem of the spectral index, which is an issue in perturbation
theories, not in background dynamics.

The spectral index obtained in our model, $n = 1 + O(m^2)$, is nearly
scale-invariant by virtue of the flatness of the potential,
which is similar to the situation in the conventional inflaton models.
In this regard, so far, the Dirac-field model does not provide
such novel features as the typical inflaton models do not have.
However, it is important to recognize that a Dirac field can drive
inflation of the universe.
As is well known, the spinor fields are indispensable not only in the
description of relativistic quantum fields, but also in the context of
supersymmetric unification of all fundamental interactions at high energy
scales.
Therefore, in the construction of realistic cosmological models containing
various matter fields and interactions between them, attention should be paid
to the behavior or properties of the spinor fields that significantly affect
the geometry of spacetime and consequently can have a central role in the
evolution of the universe.

\appendix*
\section{\label{app}Einstein-Cartan theory}

For completeness, in this appendix we briefly review the Einstein-Cartan
theory \cite{hehl}.
The Einstein-Cartan theory is a natural extension of Einstein's gravity
theory, and is one of theories that give a dynamical role to both spin and
mass of matter.

The spacetime in this theory is described by Riemann-Cartan geometry
known as a generalization of Riemann geometry to include torsion.
In the Riemann-Cartan geometry, from asymmetric affine connections
$\Gamma^\rho_{\;\;\mu\nu}$, the covariant derivative for tensors is defined by
$
\nabla_\nu V^\mu
= \partial_\nu V^\mu + \Gamma^\mu_{\;\;\rho\nu} V^\rho
$.
The connection is constrained by the metricity condition
$\nabla_{\rho}g_{\mu\nu}= 0$, which is postulated in order for a local
Minkowski structure to be guaranteed.
The curvature tensor is constructed from such connections as
\begin{equation}
R^\rho_{\ \sigma\mu\nu} \equiv
\partial_\mu \Gamma^\rho_{\;\;\sigma\nu}
- \partial_\nu \Gamma^\rho_{\;\;\sigma\mu}
+ \Gamma^\rho_{\;\;\lambda\mu} \Gamma^\lambda_{\;\;\sigma\nu}
- \Gamma^\rho_{\;\;\lambda\nu} \Gamma^\lambda_{\;\;\sigma\mu}.
\end{equation}

The difference from the Riemann geometry is that $\Gamma^\rho_{\;\;\mu\nu}$
is asymmetric.
Actually, if we demand the connection be symmetric, it can be fixed as the
well-known Riemannian connection
\begin{equation}
\tilde{\Gamma}^\rho_{\;\;\mu\nu}
= \frac{1}{2}g^{\rho\sigma}
(\partial_\nu g_{\sigma \mu} + \partial_\mu g_{\sigma \nu}
-\partial_\sigma g_{\mu \nu}).
\end{equation}
There is, however, no {\it a priori} reason that we suppose
$\Gamma^\rho_{\;\;\mu\nu}$ to be symmetric in general.
In the Riemann-Cartan geometry, the antisymmetric part is kept as
\begin{equation}
C^\rho_{\;\;\mu\nu} \equiv 2 \Gamma^\rho_{\;\;[\mu\nu]}
\equiv \Gamma^\rho_{\;\;\mu\nu} - \Gamma^\rho_{\;\;\nu\mu},
\end{equation}
which can be shown to transform as a tensor, a purely geometrical quantity.
Since the infinitesimal parallelograms do not close in this spacetime and
the closure failure is proportional to $C^\rho_{\;\;\mu\nu}$,
this tensor serves as the torsion of the spacetime.

These geometrical quantities, the curvature and the torsion, can be understood
from the local Poincar{\'e} gauge theory
\cite{hehl, hammond, utiyama, kibble, sciama} in which vierbeins $e^\mu_{\ a}$
and spin connections $\omega^{\;\;ab}_\mu$ are introduced as the gauge fields
of the theory.
In terms of the gauge fields, we can define the translational field strength
corresponding to the torsion as well as the rotational field strength
corresponding to the curvature \cite{hehl3}.

From the correspondence between a coordinate basis and a tetrad,
the vierbein satisfies
\begin{equation}
\partial_\nu e^{\mu a} + \Gamma^\mu_{\;\;\rho\nu} e^{\rho a}
+ \omega^{\;\;ab}_\nu e^\mu_{\ b} = 0.
\label{vierb.cond.}
\end{equation}
The above equation guarantees a conversion
$\nabla_\mu V^\nu = e^\nu_{\ a} D_\mu V^a$, where $D_\mu$ is the Lorentz
covariant derivative based on $\omega^{\;\; ab}_\mu$.

By virtue of the metricity condition, $\Gamma^\rho_{\;\;\mu\nu}$ can be
decomposed into the Riemannian piece and the non-Riemannian piece as
\begin{equation}
\Gamma^\rho_{\;\;\mu\nu}
= \tilde{\Gamma}^\rho_{\;\;\mu\nu}
+ \frac{1}{2}
( C^{\rho}_{\;\;\mu\nu} + C^{\ \ \>\rho}_{\mu\nu} + C^{\ \ \>\rho}_{\nu\mu}),
\label{decomp}
\end{equation}
from which it is clear that the connection reduces to the Riemannian one
when torsion vanishes.
Such decomposition holds also for the spin connection with the aid of
Eq.~(\ref{vierb.cond.}).

The basic field equations of the Einstein-Cartan theory are derived from
the simplest Lagrangian of a gravity-matter system
\begin{equation}
\mathcal{L}
( e^\mu_{\ a}, \omega^{\;\;ab}_\mu, \varphi, \partial_\mu \varphi ) =
- \frac{R}{16\pi G}
+ \mathcal{L}_m ( e^\mu_{\ a}, \varphi, D_\mu \varphi ),
\label{lagrangian}
\end{equation}
where $R \equiv e^\mu_{\ a} e^\nu_{\ b} R^{ab}_{\ \ \mu\nu}$ is the scalar
curvature in the Riemann-Cartan spacetime and $\varphi$ generically
represents matter fields minimally coupled to gravity.
Here, by making use of the relation (\ref{vierb.cond.}), not only the matter
Lagrangian $\mathcal{L}_m$, but also the Einstein-Hilbert Lagrangian can be
expressed as a function of $e^\mu_{\ a}$ and $\omega^{\;\;ab}_\mu$.

For a derivation of the field equations, it is useful to adopt the so-called
Palatini approach where $e^\mu_{\ a}$ and $\omega^{\;\;ab}_\mu$ are treated
as independent variables.
Varying the Lagrangian (\ref{lagrangian}) with respect to both, one obtains
\begin{equation}
G_\mu^{\;\; a} \equiv
R_\mu^{\;\; a} - \frac{1}{2} e_\mu^{\ a} R =
8 \pi G\> T_{\mu}^{\ a},
\label{E eq}
\end{equation}
\begin{eqnarray}
C^\mu_{\;\;ab} + 2 e^\mu_{\ [a} C^\nu_{\;\;b] \nu} &=&
- 8 \pi G i
\frac{ \partial \mathcal{L}_m }{ \partial D_\mu \varphi^A }
(S_{ab})^A_{\ B} \varphi^B
\nonumber \\
&\equiv&
- 16 \pi G S^\mu_{\;\;ab},
\label{torsion eq}
\end{eqnarray}
where $R_\mu^{\;\; a} \equiv e^\nu_{\ b} R^{ba}_{\ \ \nu\mu}$ is the
asymmetric Ricci tensor and $S_{ab}$ is the generator of the Lorentz group.
(The capital indices are spacetime or spinor indices.)

One of the resulting equations, Eq.~(\ref{E eq}), is a generalized version of
the familiar Einstein equation in the sense that the Einstein tensor
$G_\mu^{\;\; a}$ which consists of the metric-compatible and asymmetric
connection is related with the canonical energy-momentum tensor
$T_{\mu}^{\ a}$.
The other equation (\ref{torsion eq}) exhibits that the spin density
$S^\mu_{\;\;ab}$ of matter induce torsion of the spacetime.

Since the second field equation (\ref{torsion eq}) is algebraic,
we can substitute everywhere spin for torsion.
Then, according to Eq.~(\ref{decomp}), the non-Riemannian part of the
covariant derivatives in $\mathcal{L}_m$ produces an interaction between
$S^\mu_{\;\;ab}$ and $\varphi$, called a spin-interaction,
in the equation of motion for $\varphi$.

Moreover, Eq.~(\ref{decomp}) can also be applied to splitting the Einstein
tensor $G_{\mu}^{\;\;a}$ defined by Eq.~(\ref{E eq}) into the Riemannian piece
and the non-Riemannian piece.
Therefore, all the terms including the torsion in Eq.~(\ref{E eq})
can be interpreted as a spin correction to the usual, symmetric
energy-momentum tensor $\tilde{T}_{(\mu\nu)}$ through the following
expression:
\begin{equation}
\tilde{G}_{\mu\nu} =
8 \pi G \left( \tilde{T}_{(\mu\nu)} + T_{\mu\nu}^\mathrm{(spin)} \right),
\end{equation}
where $\tilde{G}_{\mu\nu}$ is the Einstein tensor composed of the Riemannian
connection $\tilde{\Gamma}^\rho_{\;\;\mu\nu}$.
The explicit form of $T_{\mu\nu}^\mathrm{(spin)}$ is given by
\begin{eqnarray}
&&
T_{\mu\nu}^\mathrm{(spin)} =
8 \pi G
\bigg[
S_\mu^{\ \rho\sigma} S_{\nu\rho\sigma}
+ 2 S^\rho_{\ \mu \rho} S^\sigma_{\ \nu \sigma}
- 4 S^{(\rho\sigma)}_{\ \ \ \ \mu} S_{(\rho\sigma)\nu}
\nonumber \\
&&
+ \frac{1}{2} g_{\mu\nu}
\left(
4 S^{(\rho\sigma)\lambda} S_{(\rho\sigma)\lambda}
- S^{\rho\sigma\lambda} S_{\rho\sigma\lambda}
- 2 S^{\sigma \rho}_{\;\ \ \sigma} S^\lambda_{\ \rho \lambda}
\right)
\bigg]
\nonumber \\
&&
+ 2 \left( \nabla_\rho - 8\pi G S^\sigma_{\ \rho\sigma} \right)
S_{(\mu\nu)}^{\ \ \ \>\rho}
+ T^\mathrm{(kin)}_{(\mu\nu)},
\end{eqnarray}
where $T^\mathrm{(kin)}_{\mu\nu}$ arises from the coupling of $\varphi$ to
$S^\mu_{\;\;ab}$ in the covariantized kinetic terms of $\mathcal{L}_m$.

The Einstein-Cartan theory is constructed in this way.
Since, as can be seen from the above equation, the spin squares contribute
to the energy-momentum in the form proportional to $G$, the predictions of
the Einstein-Cartan theory deviate from those of general relativity only when
the matter fields are coupled to gravity at high energy scales.
Therefore, it is reasonable to take the Einstein-Cartan theory
as a suitable framework for considering early stages of the universe,
in which general relativity is usually not believed to be valid.

\end{document}